\documentclass[%
 reprint,
 amsmath,amssymb,
 aps,
]{revtex4-2}
\usepackage{amssymb}
\usepackage{amsthm}
\usepackage{graphicx}
\usepackage{dcolumn}
\usepackage{bm}

\begin{document}
\title{Many-body Non-Hermitian Skin Effect  At Finite Temperatures}
\author{Kui Cao}
\affiliation{Center for Advanced Quantum Studies, Department of Physics, Beijing Normal
University, Beijing 100875, China}
\author{Qian Du}
\affiliation{Center for Advanced Quantum Studies, Department of Physics, Beijing Normal
University, Beijing 100875, China}
 
\author{Su-Peng Kou}
\thanks{Corresponding author}
\email{spkou@bnu.edu.cn}
\affiliation{Center for Advanced Quantum Studies, Department of Physics, Beijing Normal
University, Beijing 100875, China}

\begin{abstract}
In this study, we investigate the many-body non-Hermitian skin effect at finite temperatures in the thermodynamic limit. Our findings indicate an interesting correspondence between the non-Hermitian skin effect and a linear electric potential effect in this case. This correspondence leads to a unique distribution in non-Hermitian systems; particles in these many-body non-Hermitian systems do not inhabit the energy eigenstates of their single-body counterparts. As a result, the many-body non-Hermitian skin effect is significantly different from the single-body non-Hermitian skin effect. Specifically, for fermionic systems, the non-Hermitian skin effect disrupts the original phase, leading to a real-space Fermi surface. For bosonic systems, it can direct bosons to condense in corners at a decay rate that surpasses exponential, even at high temperatures. It also triggers a remarkable phase transition, resulting in spontaneous $U(1)$ symmetry breaking. Uniquely, this does not generate a Goldstone mode, presenting a deviation from traditional expectations as per the Goldstone theorem.

\end{abstract}

\maketitle

\section{introduction}

Non-Hermitian quantum systems, as a special type of open system, have been the focus of extensive research over recent years \cite{Bender98,
Bender02, Bender05, Bender07}. A certain class of translationally invariant non-Hermitian tight-binding models with non-reciprocal hopping has drawn considerable interest due to the discovery of the non-Hermitian skin effect. This effect underscores a system's dramatic divergence in physical properties between open and periodic boundary conditions, with the primary feature being the single-body eigenstates exhibit exponential localization at the boundary under open boundary conditions \cite{Yao2018,Yao20182,Xiong2018,Torres2018,Ghatak2019,Lee2019,Kunst2018,Yokomizo2019,KawabataUeda2018,SongWang2019,Longhi2019,
KZhang2020,Slager2020, YYi, OKUM20,Okuma2021,SMu2020,  Okuma2021,Roccati,ZhangYangFang2022,LiLiangWang2022}. One distinct aspect of the non-Hermitian skin effect is its manifestation in many-body systems, which exhibit a set of properties that significantly diverge from their single-body counterparts \cite{Shen Mu,E.Lee,Tliu,bose mb1,bose mb2, Zheng2023, Zhang2022,Mao2023, om2022,Dora2022}. This is in stark contrast with the Hermitian scenario, where the attributes of non-interacting many-body systems are typically mirrored by their corresponding single-body systems, driven by the simplification of the many-body system problem to a problem of single-body state filling. The non-Hermitian scenario, however, complicates this relation due to potential influences between different single-body eigenstates brought about by their non-orthogonality during the construction of many-body eigenstate. As a result, the typical correspondence between many-body and single-body systems is disrupted. This disruption is exemplified by the phenomenon of exponential localization of particles near boundaries, which is prominent in single-body non-Hermitian conditions, but is reported to be absent in fermionic many-body systems \cite{Shen Mu,E.Lee,Tliu,om2022}.

Existing research primarily approaches the many-body non-Hermitian skin effect through direct computation of many-body eigenstates.  These studies inherently focus on the zero-temperature properties of non-Hermitian systems at a small scale. However, it is crucial to recognize that realistic systems inevitably interact with their environment, leading to the unavoidable presence of finite temperatures. Moreover, many-body non-Hermitian systems at finite temperatures may manifest considerably different properties than their zero-temperature counterparts, contrasting with Hermitian situations.  In Hermitian systems, under the provision of orthogonality, a minor excitation of a limited number of particles has no influence on the single-body states of the residual unexcited particles. Therefore, variations mostly arise within the slightly excited particles contrasting the ground state, leading to substantial similarities between the ground and excited states. Furthermore, the Boltzmann distribution emphasizes  the similarities between finite-temperature and zero-temperature properties by ensuring that the ground state carries a higher weight than the excited state. However, the situation deviates considerably in non-Hermitian systems. In these systems, even minor excitations can influence the unexcited states, thereby magnifying the differences between the ground and excited states' properties. Moreover, in non-Hermitian systems, states may not necessarily align with the Boltzmann statistics, meaning the state with maximum weight may not always represent the ground state \cite{QWG3}. As such, a separate analysis becomes indispensable for finite-temperature problems.
Another noteworthy point is that condensed matter physics primarily investigates physics in the thermodynamic limit. However, existing research has yet to identify suitable methods for analyzing behaviors under such conditions, creating a gap in understanding within this field.

In this paper, we conduct an investigative study into the many-body non-Hermitian skin effect at finite temperatures, using the Hatano-Nelson model as a prototype, and focusing on the thermodynamic limit. To comprehend the behavior of the non-Hermitian system at finite temperatures, we propose a model that simulates the thermalization process of this system. Through our analysis, we find that the density matrix of the system at finite temperature could be associated with an effective Hermitian Hamiltonian, which describes a tight-binding model under the influence of a linear electric potential. Our results using the effective Hamiltonian paint a clear picture of the impact of the non-Hermitian skin effect on many-body systems. For many-body fermionic systems, it disrupts the original phase formation and prompts the emergence of a real-space Fermi surface. For many-body bosonic systems, the non-Hermitian skin effect prompts bosons to condense at corners at a rate surpassing exponential decay, even in situations with high temperatures. Additionally, we discern a significant phase transition between the condensed and the normal states within this system, a phenomenon that calls into question the validity of the Goldstone theorem in non-Hermitian systems. This observed transition triggers the spontaneous breaking of $U(1)$ symmetry but without the expected fallout of Goldstone modes.

This paper is organized in the following way.  In Section II, we delve into the statistical theory for many-body systems with the skin effect. Section III investigates the characteristics of effective models. Sections IV and V separately address the implications of the skin effect in many-body fermionic and bosonic systems at finite temperatures. We conclude the investigation and our findings in Section VI.

\section{quantum statistic
theory for Many-body  system with   Skin Effect}

In the study of many-body physics at finite temperatures, the particle distribution of the system, namely the Bose-Einstein/Fermi-Dirac distribution, plays a pivotal role as it bridges the system's single-body and many-body properties. In this section, we use the celebrated Hatano-Nelson model as an example to investigate how non-Hermitian skin effect affects the particle distribution of the system.
\subsection{ Model  }
 The Hatano-Nelson model is  described by the Hamiltonian \cite{Hatano Nelson}:
\begin{equation}
\hat{H}_{\mathrm{HN}}=J\sum_{i=1}^{L-1}(e^{-\frac{g}{2L}}a_{i+1}%
^{\dagger}a_{i}+e^{\frac{g}{2L}}a_{i}^{\dagger}a_{i+1}),
\end{equation}
where $L$ denotes the total number of lattice sites, $J$ signifies the hopping parameter, and  $g$ introduces nonreciprocity within the hopping terms, signifying the strength of the non-Hermitian skin effect. Here, $a_{i}^{\dagger}$ ($a_{i}$) encapsulates the creation (annihilation) operator for particles. For bosonic Hatano-Nelson model, it follows the commutation relation $[a_{i},a_{j}^{\dagger}]=\delta_{ij}$, while for  fermionic variant, it abides by the anti-commutation relation $\{a_{i},a_{j}^{\dagger} \}=\delta_{ij}$. Condensed matter theory customarily emphasizes the thermodynamic limit, represented by $L\rightarrow\infty$, and our investigation is explicitly geared towards this limit. Concretely, we explore cases wherein $L$ approaches infinity, all the while holding $g$ as a finite constant.
Such many-body Hatano-Nelson model can be achieved as a controlled open quantum
system \textrm{S} coupling with a Markov environment \textrm{E} 
\cite{Ueda2020,Yuto2018}. The Hamiltonian of the sub-system \textrm{S} is
$\hat{H}_{\mathrm{S}}=J\cosh({\frac{g}{2L})}\sum_{i}(a_{i+1}^{\dagger
}a_{i}+a_{i}^{\dagger}a_{i+1})$. The Lindblad operator $\mathrm{L}%
_{i}^{\mathrm{SE}}=\sqrt{2J\sinh(\frac{g}{2L}})(a_{i}+\sqrt{-1}a_{i+1})$  ($i=1,2,...,L-1$) describes the coupling between sub-system \textrm{S} and an
environment \textrm{E}. Under full-counting measurement and controlling
the number of particles on the sub-systems S, the quantum jumping processes
cause by L$_{i}^{\mathrm{SE}}$ are projected out, we get an effective
many-body Hatano-Nelson model $\hat{H}_{\mathrm{HN}}$.

We now focus on the many-body 1D  Hatano-Nelson  model at  finite temperatures. 
To achieve  many-body Hatano-Nelson model at finite temperatures, we couple the
system with a thermal bath $\mathrm{B}$  at temperature $T=\frac{1}{\beta}$. The Hamiltonian of the entire system, denoted by $\hat{H}_{tot}$, comprises three components:
\begin{equation}
\hat{H}_{tot}=\hat{H}_{\mathrm{HN}}\otimes \hat{I}_B+\hat{I}_S\otimes \hat{H}_{B}+\hat{H}_{BS},
\end{equation}
where $\hat{H}_{\mathrm{HN}}$ represents the  Hamiltonian of the Hatano-Nelson model, $\hat{H}_{B}$ corresponds to the Hamiltonian of the thermal bath, and $\hat{H}_{BS}$ denotes the coupling between the system and the thermal bath. $\hat{I}_S$ and $\hat{I}_B$ are the identity operators in the subspaces of the system and the thermal bath, respectively.  We set the coupling term as $\hat{H}_{ BS }\ =\sum
_{i}\lambda_{i}a_{i}^{\dag}a_{i}\otimes \hat{B}_{i}$. Here, $\lambda_{i}$ is a
small real coupling constant and $\hat{B}_{i}$ is an operator in thermal bath \textrm{B}. We
define that the non-Hermitian  system has the same temperature as the thermal
bath \textrm{B} at the steady state. Conversely, when the system evolves to
its steady state, its temperature approaches $T$.

\subsection{ Steady states and their effective  Hamiltonian for the Hatano-Nelson model}

\begin{figure}[ptb]
\includegraphics[clip,width=0.5\textwidth]{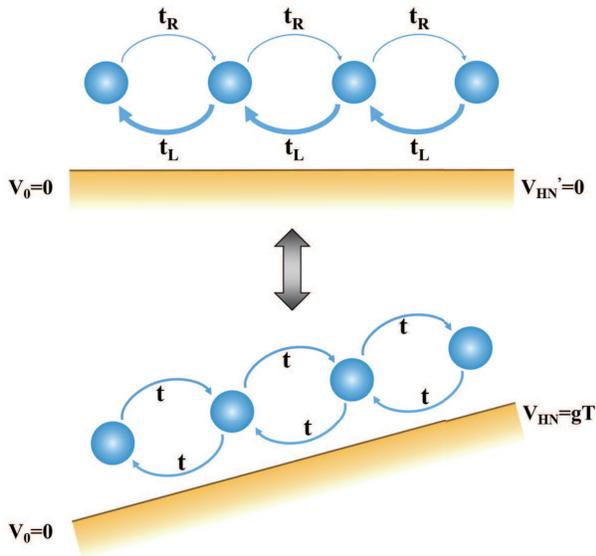}\caption{ The illustration of the correspondence between many-body non-Hermitian skin effects and linear electric potential effects at finite temperatures. } 
\end{figure}

Under our assumption, by solving the quantum master equation, the many-body Hatano-Nelson model with particle number $N$, whose steady-state density matrix $\rho _{\mathrm{T}}$ at temperature $T$, is given by (the details are provided in Appendix A)
\begin{equation}
\rho _{\mathrm{T}}= e^{-\beta \hat{H}_\mathrm{HN}  } \mathcal{T}_c ,
\end{equation}
where $\mathcal{T}_c =e^{-g\sum_{i}(\frac{i}{L}\cdot a_{i}^{\dagger}a_{i})}$.  Here, all operators are defined on the subspace with particle number $N$.
This density matrix corresponds to a non-Boltzmann distribution
\begin{equation}
\rho _{\mathrm{T}}= \sum_{n} W_n e^{-\beta E_n} \left \vert \Psi_{n}  \right \rangle \left \langle \Psi_{n} \right \vert,
\end{equation}
where
 \begin{equation*}
W_n=    (\left \langle \Psi_{n} \right \vert e^{g\sum_{i}(\frac{i}{L}\cdot a_{i}^{\dagger}a_{i})} \left \vert \Psi_{n}   \right \rangle  )^{-1} 
\end{equation*}
  is the revision of the statistical distribution of the system based on Boltzmann factor.	 $\left \vert \Psi_{n}  \right \rangle$ is the self-normalized many-body (right) eigenstate of $\hat{H}_\mathrm{HN}  $.

At zero absolute temperature, the system is in the many-body state with the lowest energy, i.e., 
\begin{equation}
\rho _{\mathrm{T}}=   \left \vert \Psi_{0}  \right \rangle \left \langle \Psi_{0} \right \vert.
\end{equation} 
This brings us back to the research of many-body physics at zero temperature \cite{Shen Mu,E.Lee,Tliu,bose mb1,bose mb2, Zheng2023, Zhang2022,Mao2023, om2022,Dora2022}.

Beyond the zero-temperature, the task of computation within the finite-temperature domain grows more involved. In such models, especially in the thermodynamic limit, the system's density matrix even at very low temperatures is typically a blend of numerous states, each bearing nontrivial weight. Consequently, the calculation of the steady-state density matrix using Eq.\,(4) evolves into a substantial challenge, yielding results that might not be intuitively comprehensible. To navigate this complexity, we propose the adoption of an effective thermal Hamiltonian, defined as follows:
\begin{equation}
e^{-\beta \hat{H}_{\mathrm{T}} }=\rho _{\mathrm{T}}.
\end{equation}
Therefore, the steady state of the many-body Hatano-Nelson model is characterized by this thermal Hamiltonian $\hat{H}_{\mathrm{T}}$.  

We analytically derive the thermal Hamiltonian given by (the detailed derivation is provided in Appendix B)
\begin{equation}
\hat{H}_{\mathrm{T}}=J_{\mathrm{eff}} \sum_{i=1}^{L-1}(a_{i+1}^{\dagger}a_{i}+h.c.)+V_{\mathrm{HN}}\sum_{i=1}^{L}\frac{i}{L}\cdot a_{i}^{\dagger}a_{i},
\end{equation}
with $J_{\mathrm{eff}}=J\frac{g }{2L \sinh (g/2L)}$
and $V_{\mathrm{HN}}=gT$. This equation contains two terms. The first term represents the Bloch Hamiltonian of a standard tight-binding model. The second term stands for a linear electric potential derived from the non-Hermitian skin effect (assuming that the particles carry a unit charge), as visualized in Fig.\,1.

During the calculations involving the effective model, we invoke the Generalized Brillouin Zone (GBZ) theory \cite{Yao2018,LeeThomale2019,YokomizoMurakami2019}, proven to be strictly applicable for bound-yet-immense systems. The theory's implementation to finite-size systems has a potential energy error $V_e \propto g/TL$ near the boundary  at the lowest order. This implies that the order of considering the thermodynamic limit and the zero-temperature limit are not interchangeable. The GBZ-induced error vanishes when taking the thermodynamic limit first, but diverges when taking the zero-temperature limit first. This observation secures our preceding conjecture and underlines the distinctive nature of finite-temperature physics from ground-state physics---the former signifies taking the thermodynamic limit first, while the latter denotes taking the zero-temperature limit first. In our ensuing discussion, we consistently adhere to taking the thermodynamic limit first, resulting in $J_{\mathrm{eff}}=J$. 

The thermal Hamiltonian proposed provides an intuitive method for characterizing and understanding the many-body non-Hermitian skin effect at finite temperatures. By employing this tool, we are able to probe into the underlying unique behaviors of such non-Hermitian systems, shedding light on the complexities of  skin phenomena in such systems in a way easily graspable and insightful.

\subsection{Particle distribution}
The quantum statistical physics of non-interacting many-body Hermitian models, denoted by many-body Hamiltonian $\hat{H}_{\mathrm{T}}$, can be mapped to the problem of state-filling in a single-body Hermitian model given by
\begin{equation}
\hat{h}_{\mathrm{T}}=J\sum_{i=1}^{L-1}(\left  \vert i\right \rangle \left \langle i+1\right \vert +h.c.)+V_{\mathrm{HN}}\sum_{i=1}^L(\frac{i}{L}\cdot \left \vert i\right \rangle \left \langle i\right \vert ).
\end{equation}
Therefore, an effective  Bose-Einstein (Fermi-Dirac) distribution  is
formed in this non-Hermitian system, i.e., when we consider the single-body operator
$\hat{\mathcal{O}}$, we have $\mathrm{tr} \  \hat{\mathcal{O}}\bar{\rho
}_{\mathrm{T}}= \mathrm{tr} \  \hat{O}\rho_{\mathrm{eff}}$, where $\bar{\rho
}_{\mathrm{T}}$ is the normalized steady-state density matrix, $\hat{O}$ is
the first quantization operator corresponding to $\hat{\mathcal{O}}$, and
\begin{equation}
\rho_{\mathrm{eff}}=\sum_{n} n ( e_{n}^{\mathrm{T}})\left \vert
e_{n}^{\mathrm{T}}\right \rangle \left \langle e_{n}^{\mathrm{T}}\right \vert ,
\end{equation}
where
\[
n (e_{n}^{\mathrm{T}})=\  \frac{1}{e^{\beta (e_{n}^{\mathrm{T}%
}-\mu)}\pm 1}.
\]
The eigenstate of $\hat{h}_{\mathrm{T}}$ with eigenvalue $e_{n}^{\mathrm{T}}$, denoted as $\left \vert e_{n}^{\mathrm{T}}\right \rangle $, is referred to as the thermal energy eigenstate. Correspondingly, $e_{n}^{\mathrm{T}}$ is referred to as the thermal energy level. For the chemical potential $\mu$, it adheres to the equation $\sum_{n}1/(e^{\beta(e_{n}^{\mathrm{T}}- \mu)} \pm 1)=N$. In this context, the negative sign refers to bosons, and the positive sign is designated for fermions. This distribution implies the presence of particles within the single-body thermal eigenstate, with the particle count in each state determined by its thermal level.
In an effort to underscore the distinction between this distribution and the conventional Bose-Einstein (Fermi-Dirac) distribution, we express this density matrix using the single-body eigenstates of $\hat{H}_{\mathrm{HN}}$ and determine
\begin{equation}
\rho_{\mathrm{eff}}=\sum_{mn}A_{mn}^{B}\left \vert \psi_{m}^{\mathrm{R}%
}\right \rangle \left \langle \psi_{n}^{\mathrm{R}}\right \vert ,
\end{equation}
where
\[
A_{mn}^{B}=\sum_{k}\frac{1}{e^{\beta (e_{k}^{\mathrm{T}}-\mu)}%
\pm 1}\left \langle \psi_{m}^{\mathrm{L}}|e_{k}^{\mathrm{T}}\right \rangle
\left \langle e_{k}^{\mathrm{T}}|\psi_{n}^{\mathrm{L}}\right \rangle .
\]
Here, $\left \vert \psi_{n}^{\mathrm{R}}\right \rangle (\left \vert \psi
_{n}^{\mathrm{L}}\right \rangle )$ is the bi-orthonormal single-body right (left) eigenstate of $\hat{H}_{\mathrm{HN}}$. 

The  ``diagonal term" $A_{nn}^{B}$, essentially informs about the average particle number in the non-Hermitian Hamiltonian eigenstate, while the ``off-diagonal" term $A_{mn}^{B}$ (for $m\neq n$), delineates the coherence between different eigenstates. Undoubtedly, this suggests that the particles no longer confine themselves to the eigenstate of the single-body non-Hermitian system. Moreover, there is a significant divergence between the thermal eigenstates and the non-Hermitian eigenstates, implying that for the many-body non-Hermitian system the behavior deviates considerably from its single-body counterpart, as depicted in Fig.\,2.

 \begin{figure}[ptb]
\includegraphics[clip,width=0.50\textwidth]{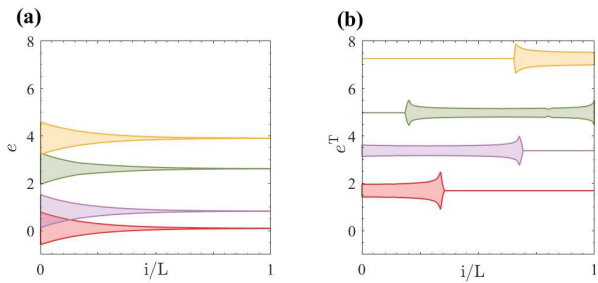}\caption{ (a) The envelopes of single-body wave functions and
corresponding  energy levels of the Hatano-Nelson model. (b) The envelopes of  thermal wave functions and
corresponding thermal  energy levels of the Hatano-Nelson
model.  The red, purple, green, and yellow bands are the envelopes of wave functions corresponding to the 500th, 1500th, 3000th, and 4500th energy levels/thermal energy levels from low to high, respectively. We set $g=5$, $J=T=1$ and $L=5000$. }%
\end{figure}

\section{non-Hermitian skin effect at finite temperatures}
In this section, we investigate the eigenvalues and eigenstates of the single-body thermal Hamiltonian Eq.\,(8) and examine how the non-Hermitian skin effect and the finite-temperature effect influence the properties of systems.
\subsection{Eigenvalues and eigenstates of single-body thermal Hamiltonian $\hat{h}_{\mathrm{T}}$}
The time-independent Schrödinger equation  
\begin{equation}
\hat{h}_{\mathrm{T}}\left \vert e_{n}^{\mathrm{T}}\right \rangle =e_{n}^{\mathrm{T}}\left \vert e_{n}^{\mathrm{T}}\right \rangle 
\end{equation}
provides the eigenvalues and eigenstates of the single-body thermal Hamiltonian $\hat{h}_{\mathrm{T}}$ (the details are provided in Appendix C).

The Schrödinger equation's solution concludes that the eigenvalue $e_{n}^\mathrm{T}$ for the $\hat{h}_{\mathrm{T}}$ is the $n$-th zero of the Lommel function $\mathcal{R}_{L,1-\delta_n}(\gamma)$. Where $\delta_n = \frac{Le_{n}^\mathrm{T}   }{gT}$, $\gamma= \frac{2JL}{gT}$.
The eigenstate $\left|e_{n}^\mathrm{T}\right\rangle$ benefits from the Wannier states $\left|i\right\rangle$ for its expansion. The expansion coefficient, $\left\langle i|e_{n}^\mathrm{T}\right\rangle$, is given by:
\begin{equation}
\left\langle i|e_{n}^\mathrm{T}\right\rangle =\frac{1}{\sqrt{\mathcal{N}}}[\mathcal{J}_{i-\delta_n}(\gamma)+k\mathcal{Y}_{i-\delta_n}(\gamma)].
\end{equation}
In this equation, $\mathcal{J}_{v}(x)$ signifies the Bessel function, whereas $\mathcal{Y}_{v}(x)$ is the Neumann function.   The term $k$ is computed as $\mathcal{J}_{-\delta_n}(\gamma)/\mathcal{Y}_{-\delta_n}(\gamma)$.
$\mathcal{N}$ marks the normalization factor. This analytical formula offers valuable insights into analyzing the  non-Hermitian skin effect at finite temperatures.

\subsection{Global phase diagram}
In varied regions, different forms of the thermal wave function give rise to distinct system phases. We introduce a dimensionless parameter $\Upsilon=\frac{J}{ V_{\mathrm{HN}} }=\frac{J }{g T }  $ to encapsulate these phases \cite{xianxing1,xianxing2,xianxing3,xianxing4}. Crossover phenomena observed at $\Upsilon \sim1$ differentiate the global phase diagram into two specific areas, referred to as the strong quantum fluctuation region ($\Upsilon>1$) and the weak quantum fluctuation region ($\Upsilon<1$). Within these regions, the eigenvalues and eigenstates associated with the single-body thermal Hamiltonian $\hat{h}_{\mathrm{T}}$ exhibit unique characteristics.

In the strong quantum fluctuation region ($\Upsilon>1$), the quantum fluctuation leads to the extension of most thermal wave functions. The dominant factor here is the kinetic term, unless the thermal energy levels are near the band edge. As a result, thermal energy levels are approximated by $2J\cos k$, with thermal wave functions labeled by quasi-momentum $k$. Only particles at the band edge are significantly affected by the effective electric potential. 

As we transition to the weak  quantum fluctuation region ($\Upsilon<1$), all thermal wave functions become localized due to the finite effective  electric potential, leading to a non-Hermitian skin effect-inspired Wannier-Stark ladder \cite{Bloch1929, Wannier1960}. The corresponding thermal wave function approximates to $\left \langle i|e_{n}^{\mathrm{T}}\right \rangle \sim \mathcal{J}_{i-n}(\gamma)$, the ratio of the local scale of the wave function to the overall system size is approximately $\Upsilon$.  These results are clearly manifesting quantum fluctuations subdued by the non-Hermitian skin effect and finite temperature effect. From the thermal energy perspective, except when near the band edge, thermal energy levels are close to $e_{n}^{\mathrm{T}}=\frac{n}{L}\cdot V_{\mathrm{HN}} $.

This detailed examination of distinct regions presents a lucid and comprehensive perspective on the non-Hermitian skin effect at finite temperatures. It cultivates a productive point of departure for exploring ramifications and potential manipulations of these non-Hermitian effects across various contexts.

\section{Many-body skin effect  in Finite-Temperature  fermionic systems}
In this section, we focus on studying the many-body skin effect in finite-temperature fermionic systems. We examine the 1D fermionic Hatano-Nelson model at half filling ($N_{F}=L/2$). Utilizing the thermal Hamiltonian $\hat{H}_{\mathrm{T}}$, we determine the spatial density of fermions as $n^F (i)=\mathrm{tr}    (a_{i}^{\dagger}a_{i}\bar{\rho
}_{\mathrm{T}})  $. The result ensues as  
\begin{equation}
n^F (i)=\sum_{n}(|\left \langle i|e_{n}^{\mathrm{T}}\right \rangle |^{2}%
\frac{1}{e^{\beta (e_{n}^{\mathrm{T}}-\mu)}+1}),
\end{equation}
as shown in Fig.\,3(a). Where $\left \vert e_{n}^{\mathrm{T}}\right \rangle $ is
the eigenstate of $\hat{h}_{\mathrm{T}}$ with eigenvalue $e_{n}^{\mathrm{T}}$. The chemical potential $\mu$ satisfies $ \sum
_{n}1/(e^{\beta(e_{n}^{\mathrm{T}}-\mu)}+1)=N_F$.

In the strong quantum fluctuation region ($\Upsilon >1$), the system resembles a fermionic tight-binding model under a mild linear potential. Within the low-temperature regime, the emergence of the skin effect is distinctly feeble, largely attenuated by quantum fluctuations.
In the domain of the weak quantum fluctuation region  ($\Upsilon<1$), the thermal energy levels is approximately expressed as $e_{n}^{\mathrm{T}}=\frac{n}{L}\cdot V_{\mathrm{HN}}$, and the coefficient $\left\langle i|e_{n}^\mathrm{T}\right\rangle$ is approximately expressed as $\delta_{ni}$. Here, we observe the emergence of an effective real-space Fermi-Dirac distribution:
\begin{align}
n^F(x) &  \sim \frac
{1}{e^{gx-\mu^{\prime}}+1},
\end{align}
where $x=i/L$. This leads to the formation of a real-space Fermi surface at large $g$.
Specifically, at high temperatures,  the system's density matrix  switches to  represent  by the pseudo-thermal Hamiltonian $\hat{H}_{\mathrm{T}}^{\prime}= \sum_{i=1}^{L}(\frac{i}{L}\cdot a_{i}^{\dagger}a_{i})$ at pseudo-temperature $T^{\prime}=\frac{L}{g}$, resulting in the emergence of an exact real-space Fermi-Dirac distribution.

 \begin{figure}[ptb]
\includegraphics[clip,width=0.47\textwidth]{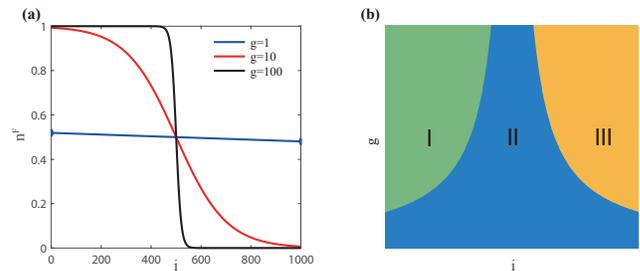}\caption{ (a) The spatial distribution
of fermions  for different $g$. We set  $L=1000$, $J=1$ and $T=0.2$. (b) A schematic representation of the impact of $g$ on the original phase at low temperatures. The label $i$ denotes lattice points. Region II represents the original phase with a particle fill rate of $1/2$. Regions I and III indicate areas where the original phase has been disrupted, with one area exhibiting a particle fill rate less than $1/2$, and the other with a rates exceeding $1/2$. }%
\end{figure}

We briefly scrutinize the scenario involving a gap, widely accounted for in the domain of condensed matter systems. In the gapped scenario, non-Hermitian skin effect can instigate a discontinuous phase transition. Starting with the Hermitian case, the potential introduced by the non-Hermitian skin effect gradually eviscerates the original formed phase, generating a unilateral particle accumulation. A paradigmatic 1D half-filling translationally invariant model serves as a viable representative for these characteristics. We concentrate predominantly on instances where the temperature $T$ is smaller than the gap $U$. Otherwise, it would make the original phase susceptible to dissolution at elevated temperatures, and lead the system to exhibit similarities with the Hatano-Nelson model.

The strength of the non-Hermitian skin effect inherent to the system is designated by $g$. For a system limited to nearest-neighbor hopping, and the potential terms can be represented either as local operators or tensor products thereof, the non-Hermitian skin effect behaves like a linear electric potential effect exhibiting   potential energy $V =   g T $ (see Appendix A for details).   To estimate the survival capacity of the original phase amidst the electric potential, we investigate a segment of the original phase of length $l$. The destruction of a nominal $\delta l$ length of the original phase at the margins by the electric potential effectuates a thermal energy modification as $\delta E_\mathrm{T}\sim (lV/L -U)\delta l $. The length of the surviving original phase can be ascribed through $\delta E_\mathrm{T}=0$, suggesting that with an added finite potential, only an original phase of length $l\sim LU/V \sim LU (g T)^{-1} $ persists, as illustrated in Fig.\,3(b).
Significantly, such phenomena exert profound influence on topological systems, insinuating that the topology of a system affected by the skin effect is essentially fragile, subject to obliteration by extremely low temperature $T\sim (g)^{-1} U$.

\section{ many-body skin effect in Finite-Temperature bosonic
 systems}
In this section, by taking the bosonic Hatano-Nelson model as an example, we investigate the many-body skin effect in finite-temperature  bosonic systems.  We
assume that the total particle number $N$ is equal to the
number of lattice sites, i.e., $N_B=L$.
\subsection{Spatial distribution  }
Using the thermal Hamiltonian $\hat{H}_{\mathrm{T}}$, we calculate the
spatial density of bosons as $n^B (i)$. The
result is obtained as
\begin{equation}
n^B (i)=\sum_{n}(|\left \langle i|e_{n}^{\mathrm{T}}\right \rangle |^{2}%
\frac{1}{e^{\beta (e_{n}^{\mathrm{T}}-\mu)}-1}),
\end{equation}
as shown in Fig.\,4. Where $\left \vert e_{n}^{\mathrm{T}}\right \rangle $ is
the eigenstate of $\hat{h}_{\mathrm{T}}$ with eigenvalue $e_{n}^{\mathrm{T}}$. 
The chemical potential $\mu$ satisfies $ \sum
_{n}1/(e^{\beta(e_{n}^{\mathrm{T}}-\mu)}-1)=N_B$. 

In particular, the spatial density of bosons at low temperatures is not approximately described by the spatial density of the ground state  ${n}^{0}(x) \sim e^{-gx}$, where $x=\frac{i}{L}$. Instead, it is approximately described by the spatial density of the
thermal \textquotedblleft ground state\textquotedblright \ \begin{equation}
{n}_{0}^{B}(i) =\frac{1}{  \mathcal{N} }[\mathcal{J}_{i-\delta_1}(\gamma)+k\mathcal{Y}_{i-\delta_1}(\gamma)]^2,
\end{equation}  i.e.,
the ground state of $\hat{h}_{\mathrm{T}}$. When $g>0$, the asymptotic behavior of this
function at $y\rightarrow \infty$ is 
\begin{equation}
{n}_{0}^{B}(x) \propto   \frac{   e^{- \frac{4}{3}y^{3/2}}}{ y^{1/2}},
\end{equation}
where $y=\frac{x}{r}$, with $r= \sqrt{\frac{J}{gTL}} $. It should be observed that we've advanced under the anticipation of the thermodynamic limit, $L \rightarrow \infty$, implying $r\rightarrow0$.   This suggests that the particles are localized at the boundary and exhibit a particle spatial density distribution decay rate that surpasses exponential decay.  

\subsection{Phase transition driven by the skin effect}

The many-body skin effect also can bring a special phase
transition. The study of condensed matter pays more attention to continuous
phase transition. In the theory of continuous phase transition, spontaneous
continuous symmetry breaking plays an important role. The traditional
continuous spontaneous symmetry breaking is accompanied by
the existence of the Goldstone mode (the Goldstone theorem). However, for the system with skin effect,
the spontaneous symmetry breaking and the existence of Goldstone mode become separate.
  We take the bosonic Hatano-Nelson model at finite temperatures as an example.
This model has a $U(1)$ symmetry, i.e.,  its  Hamiltonian invariant under the
global gauge transformation $a_{i}^{\dag}\rightarrow e^{-\sqrt{-1}\theta}a_{i}^{\dag
},a_{i}\rightarrow e^{\sqrt{-1}\theta}a_{i}$. To characterize the phase transition, we
define an order parameter $\phi_{0}=|\mathrm{tr}\ a_{0}\bar{\rho}_{\mathrm{T}}%
|^{2}=N_{0}/N_{B}$, where $a_{0}$ is the
annihilation operator for the thermal  \textquotedblleft ground
state\textquotedblright, and $N_{0}$ is the number of particles on the
thermal  \textquotedblleft ground state\textquotedblright. When a macroscopic number of particles condense on
the thermal  \textquotedblleft ground state\textquotedblright, i.e.,
$N_{0}=O(N_{B})$, we have $\phi_{0}\neq0$, at which the phase transition
occurs. On the other hand, the order parameter is zero if the state has $U(1)$
symmetry due to under the $U(1)$
gauge transformation, $\mathrm{tr}\ a_{0}\bar{\rho}_{\mathrm{T}}$ transforms as
$\mathrm{tr}\ a_{0} \bar{\rho} _{\mathrm{T}}\rightarrow e^{\sqrt{-1}\theta}\mathrm{tr}\ a_{0}%
\bar{\rho} _{\mathrm{T}}$.  Therefore, the
system exhibits spontaneous $U(1)$ symmetry breaking when this phase
transition occurs.

\begin{figure}[ptb]
\includegraphics[clip,width=0.48\textwidth]{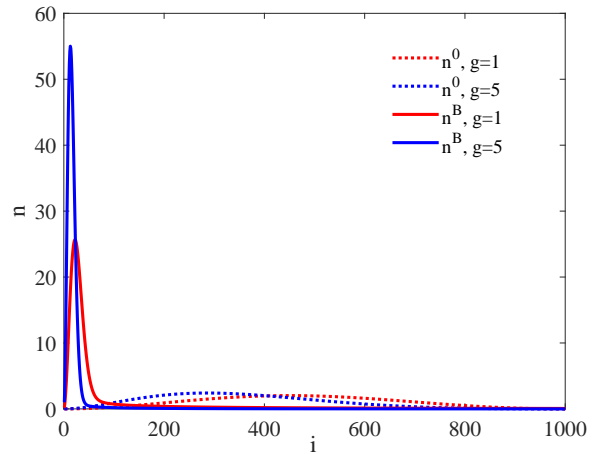}\caption{  The spatial distribution
of bosons $n^{B} $ for different $g$. We set  $L=1000$, $J=1$ and $T=0.2$. The dotted line $n^{0}$ denotes
the spatial distribution when all particles condense on the single-body ground
state of the Hatano-Nelson model. }%
\end{figure}

\begin{figure*}[ptb]
\includegraphics[clip,width=0.9\textwidth, height=0.37\textwidth]{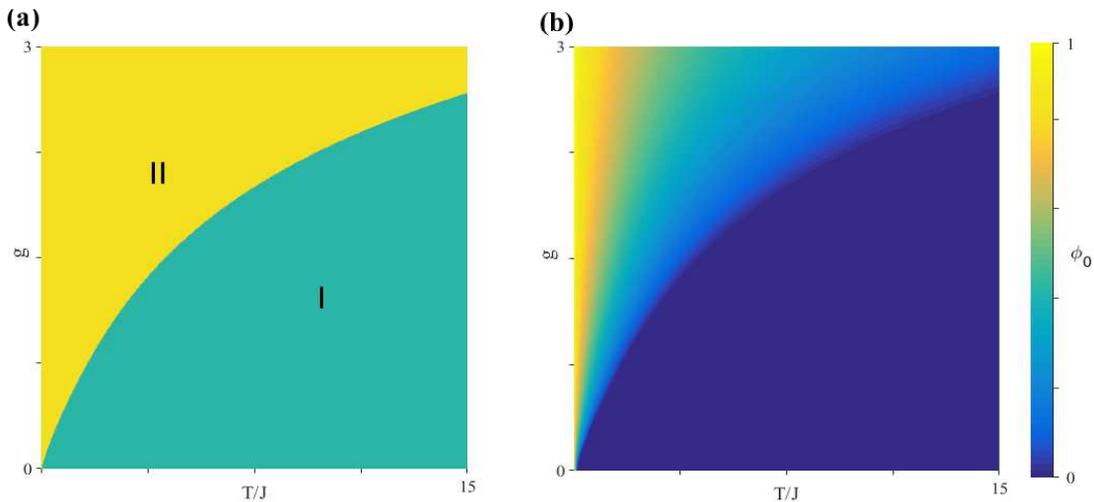}\caption{ (a) The phase 
diagram for  1D many-body bosonic Hatano-Nelson model. Region I is the normal phase, while region II has a spontaneous breaking of $U(1)$ symmetry but does not exhibit Goldstone modes. (b) The order
parameter $\phi _{0}$ for  1D many-body bosonic Hatano-Nelson model. In the region with $U(1)$ symmetry undergoing spontaneous breaking, the order parameter $\phi_{0}$ is finite. } 
\end{figure*}

The phase transition can be well defined   in the thermodynamic limit \cite{BEC0}. We estimate the values of critical
points for the phase transition. Now, the energy levels become continuous, and
the phase transition occurs at%
\begin{equation}
N_B=\int_{e_{0}^{\mathrm{T}}}^{\infty}de^{\mathrm{T}}\frac
{1}{e^{\beta(e^{\mathrm{T}}-e_{0}^{\mathrm{T}})}-1}\rho
(e^{\mathrm{T}}),
\end{equation}
where $e_{0}^{\mathrm{T}}$ is the thermal energy level with the lowest
energy, and $\rho(e^{\mathrm{T}})$ is the density of states (DOSs) derived from $\hat
{h}_{\mathrm{T}}$. The values of critical point can be obtained via
straightforward calculations.
Specificly, in the low-temperature limit, we have 
\begin{equation}
g_{c}\sim \frac{ T}{J}%
\end{equation}
or $ T_{c}\sim Jg.$  In the high-temperature limit, we have
\begin{equation}
g_{c}\sim \ln \frac{ T}{J}%
\end{equation}
or $ T_{c}\sim Je^{g}$ (the details are provided in Appendix D). The $U(1)$ symmetry  spontaneous breaking occurs in
the region of strong non-Hermitian skin effect strength $ g>g_{c} $ or low
temperature $T<T_{c}$, as seen in Fig.\,5.  This phase transition is regarded as a unique type of Bose-Einstein condensate (BEC) wherein particles condense onto the ground state of the thermal Hamiltonian. Contrary to expectations, our results show that this BEC appears in the typically forbidden area. It persists even at arbitrarily high temperatures and also breaks the Mermin-Wagner theorem \cite{BEC1,BEC2}. The emergence of this atypical  BEC  is predominantly a result of the non-Hermitian skin effect, which uniquely suppresses long-range quantum fluctuations. This suppression significantly diverges this system's behavior from standard 1D Hermitian bosonic many-body systems that possess translational symmetry. The defining characteristic in these traditional systems is the complete obliteration of BEC due to the quantum fluctuations.   It is noted that this phase transition can occur by
adjusting $g$ without changing the energy levels. Therefore,
the system can spontaneously break the $U(1)$ symmetry without generating a Goldstone mode.

\section{Conclusions}

In this study, we investigate the many-body non-Hermitian skin effect at finite temperatures. The non-Hermitian skin effect manifests significantly, akin to a linear electric potential showcasing a potential energy $V = g T$. Notably, such an equivalence is strictly valid in systems with finite temperatures in the thermodynamic limit. This strongly suggests intriguing interactions amongst finite-temperature effects, size effects, and non-Hermitian skin effects. The exploration of these interactions will be addressed in future undertakings.
Remarkably, this equivalence leads to a new distribution in such non-Hermitian systems, deviating from the conventional Bose-Einstein (Fermi-Dirac) distribution. As a result, the particles of many-body non-Hermitian systems no longer inhabit in the energy eigenstates of single-body non-Hermitian Hamiltonians. This unveils a characteristic whereby many-body systems articulate different properties from their single-body counterparts. More specifically, for fermionic systems, the non-Hermitian skin effect gradually disrupts the original phase, eventually giving birth to a real-space Fermi surface. For  bosonic systems, the non-Hermitian skin effect suppresses quantum fluctuations within the system, directing bosons to condense in corners with a decay rate that surpasses exponential. Additionally, a unique phase transition takes place between the coherent and the incoherent phases, inducing spontaneous $U(1)$ symmetry breaking that surprisingly does not generate a Goldstone mode.

The methodology demonstrated in this investigation opens up possibilities for extending this framework to a wide array of non-Hermitian systems, including high-dimensional systems and interaction systems. We anticipate further exploration in these domains in future studies.

\acknowledgments This work was supported by NSFC Grant Nos. 11974053 and 12174030. We are grateful to  Shi-Qi Zhao for helpful discussions.

\section*{ Appendix A: Density matrix of non-reciprocal models at finite temperatures}
\setcounter{equation}{0}
\renewcommand\theequation{A\arabic{equation}}
To ascertain the steady state of the non-Hermitian system with Hamiltonian $\hat{H}_{\mathrm{NH}}$, we begin by formulating the time evolution equation of the system's density matrix. Customarily, the temporal evolution of an open quantum system S is depicted by the quantum master equation \cite{Lindblad76,GKS76,BreuerPetruccione}. Generally speaking, the derivation of a quantum Markovian master equation is
performed in the interaction picture. Thus, we write the time evolution
equation of the density matrix of the system and the thermal bath  in the interaction
picture 
\begin{equation}
\frac{d}{dt}\rho^{I}_{\mathrm{S+B}}(t)=-i(\hat{V}_{I}\left(  t\right)
\rho^{I}_{\mathrm{S+B}}(t)-\rho^{I}_{\mathrm{S+B}}(t)\hat{V}_{I}^{\dag}\left(
t\right)  ). \label{2}%
\end{equation}
In the above,  $\rho^{I}_{\mathrm{S+B}}(t)=e^{i\hat{H}_{\mathrm{eff},0}t}%
\rho_{\mathrm{S+B}}(t)e^{-i\hat{H}_{\mathrm{eff},0}^{\dag}t}$  and $\hat{V}%
_{I}\left(  t\right)  =e^{i\hat{H}_{\mathrm{eff},0}t}\hat{H}_{ BS}e^{-i\hat{H}_{\mathrm{eff},0}t}$, where $\hat{V}_{I}\left(  t\right)$ is the interaction Hamiltonian in the
interaction picture. The coupling Hamiltonian is given by  $\hat{H}_{BS} = \sum_{a} \lambda_{a} \hat{C}_{a} \otimes \hat{B}_{a}$, where $\hat{C}_{a}$ and $\hat{B}_{a}$ are the operators in the subspaces of the system and the thermal bath, respectively. $\hat{H}_{\mathrm{eff},0}=\hat{H}_{\mathrm{NH}}%
\otimes \hat{I}_{\mathrm{B}}+\hat{I}_{\mathrm{S}}\otimes \hat{H}_{\mathrm{B}}$. 
Following a derivation similar to that in the Hermitian case, we obtain \cite{BreuerPetruccione,QWG3}
\begin{align}
\frac{d}{dt}\rho^{I}_{\mathrm{S}}(t)  &  =\sum_{a,b}\sum_{\omega}[\Gamma
_{ab}\left(  \omega \right)  (\hat{A}_{b}(\omega)\rho^{I}_{\mathrm{S}}%
(t)\hat{A}_{a}^{\dag}(\omega)\nonumber  \\
&  -\hat{A}_{a}(-\omega)\hat{A}_{b}(\omega)\rho^{I}_{\mathrm{S}}(t))+h.c.]. \label{10}
\end{align}
Here 
\begin{equation}
\label{7}\hat{A}_{a}(\omega)=\sum_{m}\left \vert m\right \rangle _{R}%
\left \langle m\right \vert _{L}\lambda_{a} \hat{C}_{a}\left \vert m+\omega
\right \rangle _{R}\left \langle m+\omega \right \vert _{L},
\end{equation}
with
\begin{align}
\Gamma_{ab}\left(  \omega \right)   &  =\int_{0}^{\infty}dte^{i\omega t}%
\mathrm{tr}_{\mathrm{B}}\left(  \hat{B}_{a}^{\dag}(t)\hat{B}_{b}(0)\rho
_{\mathrm{B}}^{I}\right)   
\end{align}
is the reservoir correlation function. In the above, $\left \vert m\right \rangle_{R}$ and $\left \vert m\right \rangle_{L}$ represent the bi-orthonormal right and left eigenstates of $\hat{H}_{\mathrm{NH}}$, respectively, both associated with eigenvalue $E_{m}$.  Further, $\left \vert m+\omega \right \rangle_{R/L} $ denotes the bi-orthonormal right/left eigenstate characterized by eigenvalue $E_{m}+\omega$. For simplicity, we have assumed that the energy spectrum of the system is real. $\rho
_{\mathrm{S}}^{I}$ and $\rho
_{\mathrm{B}}^{I}$ are the density matrices  of the system and the thermal bath   in the interaction picture. 

Next, we  solve the time evolution equation of the non-Hermitian system to get the steady-state solution.  For simplicity, we assume that the system  only has nearest-neighbor hopping, i.e., its Hamiltonian can be written as
\begin{equation}
\hat{H}_{\mathrm{NH}}=J\sum_{i=1}^{L-1}(t_{Ri} a_{i+1}%
^{\dagger}a_{i}+t_{Li} a_{i}^{\dagger}a_{i+1})+\hat{U},
\label{HNHG}
\end{equation}
 where $t_{Ri}$, $t_{Li}$ is the positive real hopping,  $\hat{U}$ is the potential term. We assume it can be expressed as the sum of the direct product of local Hermitian operators (common potential terms, such as real on-site potential, and coulomb interaction, all meet this condition).  The number of particles in the system is assumed to be  $N$. The models in the main text are special cases of Eq.\,(\ref{HNHG}).
 The coupling term is $\hat{H}_{ BS }=\sum_{i}%
\lambda_{i}a_{i}^{\dag}a_{i}\otimes \hat{B}_{i}$. Here, $\lambda_{i}$ is a
small real coupling constant and $\hat{B}_{i}$ is an operator in the bath \textrm{B}.
Noticed that $\left \vert m\right \rangle _{R}/\left \vert m\right \rangle _{L}$
in Eq.\,(\ref{7}) can be expressed as%
\begin{equation}
\left \vert m\right \rangle _{R}=\mathcal{\hat{S}}\left \vert m\right \rangle
_{0}, \label{8}%
\end{equation}
\begin{equation}
\left \vert m\right \rangle _{L}=(\mathcal{\hat{S}}^{-1})^{\dag}\left \vert
m\right \rangle _{0}, \label{9}%
\end{equation}
where $\mathcal{\hat{S}}=e^{ -\frac{1}{2}\sum_{j}(\sum_{i=1}^{i=j-1} \ln \frac{t_{Li}}{t_{Ri}})\cdot
a_{j}^{\dagger}a_{j} }$ 
 and $\left \vert m\right \rangle _{0}$ is the eigenstate of
$\hat{H}_{0}=\mathcal{\hat{S}}^{-1}\hat{H}_{\mathrm{HN}}\mathcal{\hat{S}%
}=\sum_{i=1}^{L-1}\sqrt{t_{Ri}t_{Li}} ( a_{i+1}%
^{\dagger}a_{i}+ a_{i}^{\dagger}a_{i+1})+\hat{U}$. We define  $\sum_{i=1}^{0} (...)=0$. According to $\hat{N}\left \vert
m\right \rangle _{R/L}=N\left \vert m\right \rangle _{R/L}$and $_{L}\left \langle
m|m\right \rangle _{R}=1$, $\left \vert m\right \rangle _{0}$ satisfy $\hat
{N}\left \vert m\right \rangle _{0}=N\left \vert m\right \rangle _{0}$ and
$_{0}\left \langle m|m\right \rangle _{0}=1$.

Substitute Eq.\,(\ref{8}), Eq.\,(\ref{9}) into Eq.\,(\ref{7}), and using Eq.\,(\ref{10}), we get
\begin{align}
\label{11}\frac{d}{dt}\rho^{I}_{\mathrm{S}}(t)  &  =\sum_{a,b}\sum_{\omega
}[\Gamma_{ab}\left(  \omega \right)  (\mathcal{\hat{S}}\hat{A}_{0,b}%
(\omega)\mathcal{\hat{S}}^{-1}\rho^{I}_{\mathrm{S}}(t)(\mathcal{\hat{S}}%
^{-1})^{\dag}\hat{A}_{0,a}^{\dag}(\omega)\mathcal{\hat{S}}^{\dag}\nonumber \\
&  -\mathcal{\hat{S}}\hat{A}_{0,a}(-\omega)\hat{A}_{0,b}(\omega)\mathcal{\hat
{S}}^{-1}\rho^{I}_{\mathrm{S}}(t))+h.c.],
\end{align}
where
\begin{equation*}
\hat{A}_{0,a}(\omega)=\sum_{m}\left \vert m\right \rangle _{0}\left \langle
m\right \vert _{0}\lambda_{a}\hat{n}_{a}\left \vert m+\omega \right \rangle
_{0}\left \langle m+\omega \right \vert _{0}.
\end{equation*}
Here $a,b=1,2,...,L$ and $\hat{n}_{a}=a_{a}^{\dag}a_{a} $.
Multiply both sides of the equal sign of Eq. (\ref{11}) by $\mathcal{\hat{S}%
}^{-1}$ to the left and $(\mathcal{\hat{S}}^{-1})^{\dag}$ to the right, we
get  
\begin{align}
&\frac{d}{dt}\mathcal{\hat{S}}^{-1}\rho^{I}_{\mathrm{S}}(t)(\mathcal{\hat{S}}^{-1})^{\dag}  \nonumber   \\ 
& =\sum_{a,b}\sum_{\omega}[\Gamma
_{ab}\left(  \omega \right)  (\hat{A}_{0,b}(\omega)\mathcal{\hat{S}}^{-1}%
\rho^{I}_{\mathrm{S}}(t)(\mathcal{\hat{S}}^{-1})^{\dag}\hat{A}_{0,a}^{\dag
}(\omega)  \nonumber  \\ 
& -\hat{A}_{0,a}(-\omega)\hat{A}_{0,b}(\omega)\mathcal{\hat{S}}^{-1}\rho^{I}_{\mathrm{S}}(t)(\mathcal{\hat{S}}^{-1})^{\dag}+h.c.]. \label{12}
\end{align}
Now, $\mathcal{\hat{S}}^{-1}\rho^{I}_{\mathrm{S}}(t)(\mathcal{\hat{S}}%
^{-1})^{\dag}$ obeys the master equation of the Hermitian system with Hamiltonian $\hat{H}_{\mathrm{0}}$. In the steady
state, the density matrix under the energy representation has only diagonal
terms. Using Eq.\,(\ref{12}), the diagonal terms of density matrix defined as
$P(n,t)=\left \langle n\right \vert _{0}\mathcal{\hat{S}}^{-1}\rho^{I}_{\mathrm{S}}(t)(\mathcal{\hat{S}}^{-1})^{\dag}\left \vert n\right \rangle
_{0}$ satisfy
\begin{equation}
\frac{d}{dt}P(n,t)=\sum_{m}[W(n|m)P(m,t)-W(m|n)P(n,t)] , \label{13}%
\end{equation}
where%
\begin{equation}
W(n|m)=\sum_{a,b}\gamma_{ab}(E_{m}-E_{n})\left \langle m\right \vert _{0}%
\lambda_{a}\hat{n}_{a}\left \vert n\right \rangle _{0}\left \langle
n\right \vert _{0}\lambda_{b}\hat{n}_{b}\left \vert m\right \rangle _{0}, 
\end{equation} 
with
\begin{align}
\gamma_{ab}(\omega) &  =\int_{-\infty}^{\infty}dte^{i\omega t} \mathrm{tr}_{\mathrm{B}}\left( \hat{B}_{a}^{\dag}(t)\hat{B}_{b}(0) \rho
_{\mathrm{B}}^{I} \right)  \nonumber  \\
&  \equiv \int_{-\infty}^{\infty}dte^{i\omega t}\left \langle \hat{B}_{a}^{\dag
}(t)\hat{B}_{b}(0)\right \rangle
\end{align}
is the real part of $2\Gamma_{ab}$. 

Using the Kubo-Martin-Schwinger  
condition $\left \langle \hat{B}_{a}^{\dag}(t)\hat{B}_{b}(0)\right \rangle
=\left \langle \hat{B}_{b}(0)\hat{B}_{a}^{\dag}(t+i\frac{1}{T})\right \rangle $,
we derive the temperature dependent behavior of $\gamma_{ab},$ i.e.,
\begin{equation}
\gamma_{ab}(-\omega)=e^{-\omega/ T}\gamma_{ba}(\omega). \label{14}
\end{equation}
When $\frac{d}{dt}P(n,t)=0$, Eq.\,(\ref{13}) and the relations Eq.\,(\ref{14})
give $W(n|m)e^{-\beta E_{m}}=W(m|n)e^{-\beta E_{n}}$ which lead to 
\begin{equation}
P(n)=const\times e^{-\beta E_{n}} 
\end{equation}
at the steady state. Then we have
\begin{equation}
\mathcal{\hat{S}}^{-1} \rho ^{I}_{\mathrm{T}}(\mathcal{\hat{S}}%
^{-1})^{\dag}=\sum_{m}\left \vert m\right \rangle _{0}e^{-\beta E_{m}%
}\left \langle m\right \vert _{0} 
\end{equation}
or
\begin{equation}
\rho ^{I}_{\mathrm{T}}=\sum_{m}\mathcal{\hat{S}}\left \vert m\right \rangle
_{0}e^{-\beta E_{m}}\left \langle m\right \vert _{0}\mathcal{\hat{S}}^{\dag},
\end{equation}
where $\rho ^{I}_{\mathrm{T}}$ is the steady state in the interaction
picture, which is also the steady state
\begin{equation}
\rho_{\mathrm{T}}=\sum_{m}\mathcal{\hat{S}}\left \vert m\right \rangle
_{0}e^{-\beta E_{m}}\left \langle m\right \vert _{0}\mathcal{\hat{S}}^{\dag
}%
\end{equation}
in the Schrödinger picture. Therefore, we have \begin{equation} \rho _{\mathrm{T}}= \mathcal{\hat{S}}e^{-\beta \hat{H}_{0} }\mathcal{\hat{S}}^{\dagger}=e^{-\beta \hat{H}_\mathrm{NH}  } \mathcal{T}_c,\end{equation}
here  $\mathcal{T}_c=e^{ -\sum_{j} (\sum_{i=1}^{i=j-1}  \ln \frac{t_{Li}}{t_{Ri}} )\cdot
a_{j}^{\dagger}a_{j} }$.

For translationally invariant models, by defining $g= L \ln \frac{t_{Li}}{t_{Ri}} $, we can achieve $\mathcal{T}_c =e^{-g\sum_{i}(\frac{i-1}{L}\cdot a_{i}^{\dagger}a_{i})}$, which is equivalent to $\mathcal{T}_c =e^{-g\sum_{i}(\frac{i}{L}\cdot a_{i}^{\dagger}a_{i})}$. Based on the definition   $e^{-\beta \hat{H}_{\mathrm{T}} }=\rho _{\mathrm{T}}$, and utilizing the  Baker-Campbell-Hausdorff (BCH) formula, we can derive the thermal Hamiltonian  $\hat{H}_{\mathrm{T}}$ as
\begin{equation}
\hat{H}_{\mathrm{T}}= \hat{H}_0 +V\sum_{i=1}^{L}\frac{i}{L}\cdot a_{i}^{\dagger}a_{i}+...,
\end{equation}
with $V=gT$. In the thermodynamic limit, contributions from higher-order terms tend to zero. Hence, in such a limit and at finite temperatures, it follows that the many-body skin effect becomes equivalent to a linear electric potential effect.

\section*{ Appendix B:  thermal Hamiltonian for   Hatano-Nelson model}
\setcounter{equation}{0}
 \renewcommand\theequation{B\arabic{equation}}


  To calculate the thermal Hamiltonian for the Hatano-Nelson model, we first rewrite the density matrix as
\begin{equation}
\rho _{\mathrm{T}}= e^{-\beta \hat{H}_\mathrm{HN}  } \mathcal{T}_c=e^{\hat{A}}e^{\hat{B}+\hat{B}^{\dag}}e^{\hat{A}%
},
\end{equation}
where
\begin{equation}
\hat{A}=-\frac{g}{2L}\sum_{i=1}^{L}(ia_{i}^{\dagger}a_{i}),\text{ }%
\hat{B}=-\beta J\sum_{i=1}^{L-1}(a_{i+1}^{\dagger}a_{i}).
\end{equation}
We have
\begin{equation}
\lbrack \hat{A},\hat{B}]=-\frac{g}{2L}\hat{B},\text{ }[\hat{A},\hat
{B}^{\dag}]=-\frac{g}{2L}\hat{B}^{\dag}.
\end{equation}
In the following parts, we set $-\frac{g}{2L}=a$. In order to get a
simple analytic result, we use the GBZ theory.  Now, we have
\begin{equation}
\hat{B}=-\beta Je^{\sqrt{-1}k} \hat{n}_{k},\text{ }\hat{B}^{\dag}=-\beta Je^{-\sqrt{-1}k}\hat{n}_{k},
\end{equation}
where $k$ is quasi-momentum and  $\hat{n}_{k}$ is the number operator for the state with quasi-momentum $k$.  In particular, we have $[\hat{B},\hat{B}^{\dag}]=0$. As
a result, the density matrix $\rho _{\mathrm{T}}$ is transformed into
\begin{equation}
\rho _{\mathrm{T}}=e^{\hat{A}}e^{\hat{B}+\hat{B}^{\dag}}e^{\hat{A}}=e^{\hat{A}}e^{\hat{B}%
}e^{\hat{B}^{\dag}}e^{\hat{A}}.
\end{equation}

Next,  using the formula $e^{\hat
{C}+\hat{D}}=e^{\hat{C}}e^{f(c)\hat{D}}$, where $f(c)=c/(1-e^{-c})$ with  
$\hat{C}$, $\hat{D}$ satisfy $[\hat{C},\hat{D}]=c\hat{D},$ we can get
\begin{equation}
e^{\hat{C}}e^{\hat{D}}=e^{\hat{C}+g(c)\hat{D}},
\end{equation}
where $g(c)=f(c)^{-1}.$ Here, we have used $[\hat{C},(\xi \hat{D})]=c(\xi
\hat{D})$, where $\xi$ is a arbitrary constant. As a result, we have%
\begin{equation}
e^{\hat{A}}e^{\hat{B}}=e^{\hat{A}+g(a)\hat{B}}%
\end{equation}
and
\begin{equation}
e^{\hat{B}^{\dag}}e^{\hat{A}}=[e^{\hat{A}}e^{\hat{B}}]^{\dag}=e^{\hat
{A}+g(a)\hat{B}^{\dag}}.
\end{equation}
Therefore, we transform the density matrix
\begin{equation}
\rho _{\mathrm{T}}=e^{\hat{A}}e^{\hat{B}}e^{\hat{B}^{\dag}}e^{\hat{A}}%
\end{equation}
to
\begin{equation}
\rho _{\mathrm{T}}=e^{\hat{A}+g(a)\hat{B}}e^{\hat{A}+g(a)\hat{B}^{\dag}}. \label{101}
\end{equation}
Noticing that if we start with another expression of $\rho _{\mathrm{T}}$:
\begin{equation}
\rho _{\mathrm{T}}=e^{\hat{A}}e^{\hat{B}+\hat{B}^{\dag}}e^{\hat{A}}=e^{\hat{A}}e^{\hat
{B}^{\dag}}e^{\hat{B}}e^{\hat{A}},
\end{equation}
we have 
\begin{equation}
\rho _{\mathrm{T}}=e^{\hat{A}+g(-a)\hat{B}^{\dag}}e^{\hat{A}+g(-a)\hat{B}}.  \label{102}
\end{equation}

Finally, by using BCH formula on Eq.\,(\ref{101}) and Eq.\,(\ref{102}), and compare the two results, we find that 
\begin{align}
\rho _{\mathrm{T}} &  =e^{2\hat{A}+g(-a)(\hat{B}+\hat{B}^{\dag})+g(-a)\frac{1-g(-a)/g(a)}%
{1+g(-a)/g(a)}(\hat{B}+\hat{B}^{\dag})} \nonumber \\
&  =e^{2\hat{A}+\frac{2g(a)g(-a)}{g(a)+g(-a)}(\hat{B}+\hat{B}^{\dag}%
)},
\end{align}
i.e., the density matrix becomes
\begin{equation}
\rho _{\mathrm{T}}    =\exp \{-\beta \sum_{i}J_{\mathrm{eff}} (a_{i+1}^{\dagger}a_{i}%
+h.c.)-g\sum_{i}(\frac{i}{L} \cdot a_{i}^{\dagger}a_{i})\}
\end{equation}
where
\begin{equation}
J_{\mathrm{eff}}=J\frac{2g(a)g(-a)}{g(a)+g(-a)}=\frac{g }{2L \sinh (g/2L)}.
\end{equation}
Therefore, the analytical formula of thermal Hamiltonian in the thermodynamic limit is obtained as  
\begin{align}
\hat{H}_{\mathrm{T}} &  =-\frac{1}{\beta }\ln \rho _{\mathrm{T}}\nonumber \\
&  =J_{\mathrm{eff}}\sum_{i}(a_{i+1}^{\dagger}a_{i}+h.c.)+V_{\mathrm{HN}}\sum_{i}%
(\frac{i}{L}\cdot a_{i}^{\dagger}a_{i} ),
\end{align}
with $J_{\mathrm{eff}}=J\frac{g }{2L \sinh (g/2L)}$
and $V_{\mathrm{HN}}=gT$.

\section*{ Appendix C: Solving for Eigenvalues and eigenstates of Single-Body thermal Hamiltonian $\hat{h}_{\mathrm{T}}$}
\renewcommand\theequation{C\arabic{equation}}
 Expand the single-body thermal Hamiltonian and the eigenstate with bases
$\left \vert i\right \rangle$, we can get a differential equation
\begin{equation}
J[\psi_{n}(i-1)+\psi_{n}(i+1)]=(e_{n}^{\mathrm{T}}%
-i\varepsilon_{\mathrm{HN}})\psi_{n}(i),\  \  \ i\in \mathbb{Z}%
\end{equation}
with the open boundary condition
\begin{equation}
\psi_{n}(0)=\psi_{n}(L+1)=0.
\end{equation}
Here, $\psi_{n}(i)\equiv \left \langle i|e_{n}^{\mathrm{T}}\right \rangle $ is
the expansion coefficient of $\left \vert e_{n}^{\mathrm{T}}\right \rangle $, $\varepsilon_{\mathrm{HN}}= V_{\mathrm{HN}}/L$.
The solution of the Bessel equation satisfies the recurrence relation
\begin{equation}
\mathcal{Z}_{\nu+1}(x)+\mathcal{Z}_{\nu-1}(x)=2\nu/x\; \mathcal{Z}_{\nu}(x).
\end{equation}
As a result, we have
\begin{equation}
\psi_{n}(i)=A\; \mathcal{J}_{i-e_{n}^{\mathrm{T}}/\varepsilon_{\mathrm{HN}}%
}(\frac{2J}{\varepsilon_{\mathrm{HN}}})+B\; \mathcal{Y}%
_{i-e_{n}^{\mathrm{T}}/\varepsilon_{\mathrm{HN}}}(\frac{2J%
}{\varepsilon_{\mathrm{HN}}}),
\end{equation}
where $\mathcal{J}_{v}(x)$ is the Bessel function, $\mathcal{Y}_{v}(x)$ is
Neumann function, and $A$, $B$ are some constant that is determined by
boundary conditions and normalization.

Next, we calculate $A$ and $B$ from boundary conditions. Using boundary
conditions, we have
\begin{equation}
\label{20}A\; \mathcal{J}_{-e_{n}^{\mathrm{T}}/\varepsilon_{\mathrm{HN}}%
}(\frac{2J}{\varepsilon_{\mathrm{HN}}})+B\; \mathcal{Y}%
_{-e_{n}^{\mathrm{T}}/\varepsilon_{\mathrm{HN}}}(\frac{2J%
}{\varepsilon_{\mathrm{HN}}})=0,
\end{equation}
and%
\begin{equation}
A\; \mathcal{J}_{L+1-e_{n}^{\mathrm{T}}/\varepsilon_{\mathrm{HN}}}%
(\frac{2J}{\varepsilon_{\mathrm{HN}}})+B\; \mathcal{Y}%
_{L+1-e_{n}^{\mathrm{T}}/\varepsilon_{\mathrm{HN}}}(\frac{2J%
}{\varepsilon_{\mathrm{HN}}})=0.
\end{equation}
Regard the above equations as homogeneous linear equations about $A$ and $B$,
the conditions for the existence of nonzero solutions for energy levels is
\begin{equation}%
\begin{vmatrix}
\mathcal{J}_{-e_{n}^{\mathrm{T}}/\varepsilon_{\mathrm{HN}}}(\frac
{2J}{\varepsilon_{\mathrm{HN}}}) & \mathcal{Y}_{-e_{n}%
^{\mathrm{T}}/\varepsilon_{\mathrm{HN}}}(\frac{2J}%
{\varepsilon_{\mathrm{HN}}})\\
\mathcal{J}_{L+1-e_{n}^{\mathrm{T}}/\varepsilon_{\mathrm{HN}}}(\frac
{2J}{\varepsilon_{\mathrm{HN}}}) & \mathcal{Y}_{L+1-e_{n}%
^{\mathrm{T}}/\varepsilon_{\mathrm{HN}}}(2\frac{J}%
{\varepsilon_{\mathrm{HN}}})
\end{vmatrix}
=0,
\end{equation}
i.e.,  \begin{widetext}
\begin{equation}
\mathcal{J}_{-e_{n}^{\mathrm{T}}/\varepsilon_{\mathrm{HN}}}%
(\frac{2J}{\varepsilon_{\mathrm{HN}}})\mathcal{Y}%
_{L+1-e_{n}^{\mathrm{T}}/\varepsilon_{\mathrm{HN}}}(\frac
{2J}{\varepsilon_{\mathrm{HN}}})-\mathcal{Y}%
_{-e_{n}^{\mathrm{T}}/\varepsilon_{\mathrm{HN}}}(\frac{2J }{\varepsilon_{\mathrm{HN}}})\mathcal{J}_{L+1-e_{n}^{\mathrm{T}}%
/\varepsilon_{\mathrm{HN}}}(\frac{2J}{ \varepsilon_{\mathrm{HN}}})=0.\end{equation}
\end{widetext}
Using the Lommel polynomial,
\begin{equation}
\mathcal{R}_{N,\nu}(x)=\frac{1}{2}\pi x[Y_{N+\nu}(x)\mathcal{J}_{\nu
-1}(x)-\mathcal{J}_{N+\nu}(x)Y_{\nu-1}(x)],
\end{equation}
the energy levels $e_{n}^{\mathrm{T}}$ are obtained by solving the following
equation
\begin{equation}
\mathcal{R}_{L,1-e_{n}^{\mathrm{T}}/\varepsilon_{\mathrm{HN}}}(\frac
{2J}{\varepsilon_{\mathrm{HN}}})=0,
\end{equation}
or
\begin{equation}
\mathcal{R}_{L,1-\delta_n}(\gamma)=0.
\end{equation}
where  $\delta_n = \frac{e_{n}^{\mathrm{T}}}{\varepsilon_{\mathrm{HN}}}= \frac{Le_{n}^\mathrm{T}   }{gT}$, $\gamma= \frac
{2J}{\varepsilon_{\mathrm{HN}}}= \frac{2JL}{gT}$.
Substitute the solved energy level $e_{n}^{\mathrm{T}}$ into Eq.\,(\ref{20}),
we have
\begin{equation}
B/A=\mathcal{J}_{-\delta_n}(\gamma)/\mathcal{Y}_{-\delta_n}(\gamma).
\end{equation}
Then the corresponding wave function $\psi_{n}(i)$ is obtained as
\begin{equation}
\psi_{n}(i)=\frac{1}{\sqrt{\mathcal{N}}}[\mathcal{J}_{i-\delta_n}(\gamma)+k\mathcal{Y}_{i-\delta_n}(\gamma)],
\end{equation}
where
\begin{equation}
k=B/A=\mathcal{J}_{-\delta_n}(\gamma)/\mathcal{Y}_{-\delta_n}(\gamma),
\end{equation}
$\mathcal{N}$ is the normalization factor determined by
\begin{equation}
\mathcal{N}=\sum_{i=1}^{L}\left \vert \mathcal{J}_{i-\delta_n}(\gamma)+k\; \mathcal{Y}_{i-\delta_n
}(\gamma)\right \vert ^{2}.
\end{equation}

\section*{ Appendix D: ESTIMATION OF CRITICAL POINTS  FOR
 bosonic Hatano-Nelson model}
\setcounter{equation}{0}
\renewcommand\theequation{D\arabic{equation}} 
 \subsection{Density of states}
We summarize the approximate expression of the energy spectrum of $\hat{h}_{\mathrm{T}}$ in some limited
cases in this subsection \cite{xianxing1,xianxing2,xianxing3,xianxing4}. We have set the ground state
energy to zero. To characterize the effective model, we have introduced a dimensionless parameter:
$\Upsilon= \frac{J}{gT}$. In the strong quantum fluctuation  limit $\Upsilon \gg 1$, the
energy levels are obtained as
\begin{equation}
e_{{n}}^{\mathrm{T}}=%
\begin{cases}
2J(1- \cos  \frac{2\pi n}{L}) &  \ n\in \mathbb{N}_{+},e_{{}}%
^{\mathrm{T}}\gg V_{\mathrm{HN}} \\
J(JL/V_{\mathrm{HN}})^{-\frac
{2}{3}}[\frac{3\pi(n+\frac{3}{4}))}{2}]^{\frac{2}{3}} & n\in \mathbb{N}%
_{+},e_{{}}^{\mathrm{T}}\ll V_{\mathrm{HN}}%
\end{cases}
.
\end{equation}
In the weak quantum fluctuation  limit $\Upsilon \ll 1$, the energy levels are
obtained as%
\begin{equation}
e_n^{\mathrm{T}}=%
\begin{cases}
V_{\mathrm{HN}} \frac{n}{L} & n\in \mathbb{N}_{+},e_{{}}^{\mathrm{T}%
}\gg J\\
J(JL/V_{\mathrm{HN}})^{-\frac
{2}{3}}[\frac{3\pi(n+\frac{3}{4}))}{2}]^{\frac{2}{3}} & n\in \mathbb{N}%
_{+},e_{{}}^{\mathrm{T}}\ll J%
\end{cases}
.
\end{equation}
The   DOSs  is defined as
\begin{equation}
\rho(e_{{}}^{\mathrm{T}})=\frac{dn}{de_{{}}^{\mathrm{T}}}.
\end{equation}
In the thermodynamic limit, the DOSs of the thermal
Hamiltonian $\hat{h}_{\mathrm{T}}$ is obtained: 

In the strong quantum fluctuation limit, we have
\begin{equation}
\rho(e_{{}}^{\mathrm{T}})=%
\begin{cases}
L \frac{1}{\pi \sqrt{J }V_{\mathrm{HN}} }\sqrt{e_{{}}^{\mathrm{T}}} & e_{{}}^{\mathrm{T}}\ll V_{\mathrm{HN}} \\
L\frac{1}{2\pi \sqrt{J  }}(e_{{}}^{\mathrm{T}})^{-\frac{1}{2}} &
e_{{}}^{\mathrm{T}}\gg V_{\mathrm{HN}} 
\end{cases}
.
\end{equation}

In the weak quantum fluctuation limit, we have
\begin{equation}
\rho(e_{{}}^{\mathrm{T}})=%
\begin{cases}
L\frac{1}{\pi}\frac{1}{\sqrt{J }V_{\mathrm{HN}} }\sqrt{e_{{}}^{\mathrm{T}}} & e_{{}}^{\mathrm{T}}\ll J  \\
L\frac{1}{V_{\mathrm{HN}} } &   V_{\mathrm{HN} }
 >e_{{}}^{\mathrm{T}}\gg J \\
0 & e_{{}}^{\mathrm{T}}\geq V_{\mathrm{HN}} 
\end{cases}
.
\end{equation}

\subsection{Results in the low-temperature limit}

Let's estimate the critical point of the phase transitions between the
BEC and the normal state in the low-temperature limit. 
The phase transition occurs at $N_B= \int_{e_{0}^{\mathrm{T}}}^{\infty}de^{\mathrm{T}}\frac{1}{e^{\beta (e^{\mathrm{T}}-e_{0}^{\mathrm{T}})}-1}\rho(e_{{}%
}^{\mathrm{T}}),$ where $N_B$ is the number of particles, and we set $N_B=L$.
$e_{0}^{\mathrm{T}}$ is the minimum eigenvalue of thermal Hamiltonian
$\hat{h}_{\mathrm{T}}$, and the DOSs $\rho(e_{{}}^{\mathrm{T}})$
is approximate to (it turns out that the critical point in the low-temperature
limit occurs in the strong quantum fluctuation  limit. If we use the DOSs
of the weak quantum fluctuation limit to calculate, we will get inconsistent
results).
\begin{equation}
\rho(e_{{}}^{\mathrm{T}})=%
\begin{cases}
L \frac{1}{\pi \sqrt{J }V_{\mathrm{HN}}%
}\sqrt{e_{{}}^{\mathrm{T}}} & \Lambda_{l}^{1}>e_{{}}^{\mathrm{T}}\geq0\\
L\frac{1}{2\pi \sqrt{J }}({e^{\mathrm{T}}})^{-{\frac{1}{2}}} &
e^{\mathrm{T}}\geq \Lambda_{l}^{1}%
\end{cases}
,
\end{equation}
where $\Lambda_{l}^{1}$ is a cutoff and can be written as $\alpha_{l}^{1}%
 gT$, $\alpha_{l}^{1}$ is a real constant of the
order of magnitude around $1$.

From the fact of
\begin{align}
N_B/L    =\frac{1}{\pi}[\frac{1}{\sqrt{J }V_{\mathrm{HN}} }\int_{0}^{\Lambda_{l}^{1}}de^{\mathrm{T}}\frac{\sqrt{e_{{}}^{\mathrm{T}}}%
}{e^{\beta e_{{}}^{\mathrm{T}}}-1} +\frac{1}{\sqrt{2J }}\int_{\Lambda_{l}^{1}}^{\infty}de_{{}%
}^{\mathrm{T}}\frac{({e^{\mathrm{T}}})^{-{\frac{1}{2}}}}{e^{\beta e_{{}%
}^{\mathrm{T}}}-1}],
\end{align}
we have
\begin{align}
N_B/L  =\frac{2}{\pi}(\sqrt{\alpha_{l}^{1}}+\sqrt{\frac{1}{\alpha_{l}^{1}}}%
)\sqrt{\frac{ T}{J g}}.
\end{align}
As a result, the  critical point of the phase transition is
\begin{equation}
 T_{c}\sim Jg.
\end{equation}

\subsection{Results in the high-temperature limit}

It turns out that the critical point in the high-temperature limit occurs in the weak quantum fluctuation limit. By using the approximate of the DOSs,
\begin{equation}
\rho(e_{{}}^{\mathrm{T}})=%
\begin{cases}
L\frac{1}{\pi \sqrt{J }V_{\mathrm{HN}}%
}\sqrt{e_{{}}^{\mathrm{T}}} & e_{{}}^{\mathrm{T}}<\Lambda_{h}^{1}\\
L\frac{1}{V_{\mathrm{HN}}} & V_{\mathrm{HN}}>e_{{}}^{\mathrm{T}}\geq \Lambda_{h}^{1}\\
0 & e_{{}}^{\mathrm{T}}\geq V_{\mathrm{HN}} 
\end{cases}
,
\end{equation}
($\Lambda_{ht}$ is a cutoff and can be written as $\alpha_{h}%
^{1}J $, $\alpha_{h}^{1}$ is a real constant of the order of
magnitude around $1$), we have
\begin{align}
N_B/L  &  =\frac{1}{\pi}[\frac{1}{\sqrt{J }V_{\mathrm{HN}} }\int_{0}^{\Lambda_{h}^{1}}de_{{}}^{\mathrm{T}}\frac
{\sqrt{e ^{\mathrm{T}}}}{e^{\beta e ^{\mathrm{T}}}-1} +\int_{\Lambda_{h}^{1}}^{V_{\mathrm{HN}} }de_{{}%
}^{\mathrm{T}}\frac{1}{e^{\beta e_{{}}^{\mathrm{T}}}-1}\frac{1}%
{V_{\mathrm{HN}} }.
\end{align}
We expect that in the high-temperature limit, $g$ may
be large, and the first term can be ignored. Integral the second term, we get
\begin{equation}
N_B/L=\frac{1}{g}\ln \frac{1-e^{-g}}{1-e^{-\alpha_{h}^{1}J / T_{c}}}=1
\end{equation}
that gives%
\begin{align}
g  &  =\ln(\frac{1}{2(1-e^{-\alpha_{h}%
^{1}J / T_{c}})}[1+\sqrt{1-4(1-e^{-\alpha_{h}^{1}%
J / T_{c}})}])  \sim 
\ln \frac{ T_{c}}{J}+O(1)
\end{align}
or
\begin{equation}
 T_{c}\sim Je^{g}.
\end{equation}

\end{document}